\documentclass[apj]{emulateapj}
\usepackage{refcount}
\usepackage{amsmath}
\usepackage{amssymb}
\usepackage{hyperref}
\usepackage{euscript}
\usepackage{algorithm2e,algorithmic}
\usepackage{bbold}

\usepackage{paralist}

\def\gtorder{\mathrel{\raise.3ex\hbox{$>$}\mkern-14mu
             \lower0.6ex\hbox{$\sim$}}}
\def\ltorder{\mathrel{\raise.3ex\hbox{$<$}\mkern-14mu
             \lower0.6ex\hbox{$\sim$}}}

\slugcomment{Draft of \today}
\shortauthors{Ofek}

\begin{document}

\title{Towards the measurement of the mass of isolated Neutron Stars - Prediction of future astrometric microlensing events by Pulsars}
\author{Eran O. Ofek\altaffilmark{1}}

\altaffiltext{1}{Benoziyo Center for Astrophysics, Weizmann Institute
  of Science, 76100 Rehovot, Israel}

\begin{abstract}

The mass of single neutron stars (NSs) can be measured using
astrometric microlensing events.
In such events, the center-of-light motion of a star lensed by a NS
will deviate from the expected non-lensed motion and this deviation can be used
to measure the mass of the NS.
I search for future conjunctions between pulsars, with
measured proper motion,
and stars in the GAIA-DR2 catalog.
I identified one candidate event of a star that will possibly be lensed by a pulsar
during the next ten years in which the expected light deflection
of the background star will deviate from the non-lensed motion
by more than 50\,$\mu$as.
%
%PSR\,J$1856-3754$ will pass $\cong0.6''$ from a $20.4$\,$G$\,magnitude background star in 2027.8,
%while 
Given the position and proper motion of PSR\,J$0846-3533$,
it will possibly pass $\sim0.2''$ from a 19.0\,$G$\,magnitude background star in 2022.9.
Further assuming a 1.4\,M$_{\odot}$ NS,
the expected maximum deviation of the background star images
from the uniform-rate plus parallax motion will be
%115\,$\mu$as and
91\,$\mu$as.
%for PSR\,J$1856-3754$ and PSR\,J$0846-3533$, respectively.
%
This pulsar position has a relatively large uncertainty
and therefore additional observations are required in order to verify this event.
I briefly discuss the opposite case, in which a pulsar is being lensed by a star.
Such events can be used to measure the stellar mass via pulsar timing measurements.
I do~not find good candidates for such events with predicted variations
in the pulsar period derivative ($\dot{P}$), divided by 1\,s,
exceeding $10^{-20}$\,s$^{-1}$.
Since only about 10\% of the known pulsars have measured
proper motions, there is potential for an increase
in the number of predicted pulsar lensing events.

\end{abstract}

\keywords{astrometry --- 
gravitational lensing: micro ---
stars: neutron}
%methods: statistical ---
%techniques: image processing ---
%techniques: photometric}

\section{Introduction}
\label{sec:Introduction}

The mass and population mass-range of Neutron Stars (NSs) are fundamental properties related
to their formation and evolution and to the equation of state of nuclear matter.
So far, we have only obtained accurate mass measurements for NSs in binary systems
(e.g., Kramer \& Stairs 2008).
There is some evidence that there is more than one channel to form
a NS (e.g., Beniamini \& Piran 2016).
Therefore, measuring masses of single NS is of great importance.

Paczy{\'n}ski (1995; 1998), Miralda-Escude (1996), and Gould (2000)
have suggested to measure stellar masses via the detection of astrometric
microlensing events.
In such events, a source is lensed by a stellar mass object
in our galaxy, and the center-of-light of the source images
will deviate from a uniform-rate proper motion.
This deviation could be used to measure the mass of the lensing star.
Sahu et al. (1998) and Salim \& Gould (2000)
made some predictions
for future astrometric microlensing events.
Harding et al. (2018) estimated the astrometric microlensing rate for known stellar remnants,
and McGill et al. (2018) used
the GAIA-TGAS catalog (Lindegren et al. 2016) to make predictions for astrometric microlensing events.
Furthermore, Bramich et al. (2018) and Mustill et al. (2018) made some predictions for microlensing
events in the next ten and twenty years, respectively, based on the GAIA-DR2 catalog.
Finally, Lu et al. (2016) and Kains et al. (2017)
presented some onging efforts to measure astrometric microlensing
events focusing on identifying single stellar-mass black holes in
our galaxy,
while Sahu et al. (2017) presented the first measurements of a white dwarf mass
based on an astrometric microlensing event.

Here, I search for future close angular conjunctions between
pulsars with a known proper motion and stars in the GAIA-DR2 catalog
(Gaia Collaboration et al. 2016; Gaia Collaboration et al. 2018).
I find one candidate event in the nearby future, in which a pulsar will pass with
a small angular separation from a background star.
The impact parameter of this close angular passage is
hundreds of times the Einstein radius of the lens.
Therefore, this event will not result in a classical
microlensing -- i.e., events that have a detectable magnification of the background star.
However, this event may produce small astrometric shift in the position
of the background star relative to the linear motion at a constant
angular speed expected from the proper motion component
and the periodic variation expected from the parallax.
Future measurements of such astrometric microlenisng events
may enable the first mass measurements of a single NS.

In \S\ref{sec:search}, I describe the search for astrometric microlensing events involving pulsars,
and the candidate is listed in \S\ref{sec:Cand}.
The results are discussed in \S\ref{sec:Disc}.

\section{The search}
\label{sec:search}

I selected all the pulsars listed in the Australia Telescope National Facility (ATNF) pulsar
database\footnote{http://www.atnf.csiro.au/people/pulsar/psrcat/}
(version 1.58 of date 2018 May 10) that have proper motion measurements.
The declination of PSR\,J$1856-3754$ in the ATNF catalog was erroneous\footnote{The erroneous declination
led to the detection of a spurious event that was listed in an earlier version of this paper.}.
and I use the correct declination from Walter \& Matthews (1997).

\begin{figure}
\centerline{\includegraphics[width=8cm]{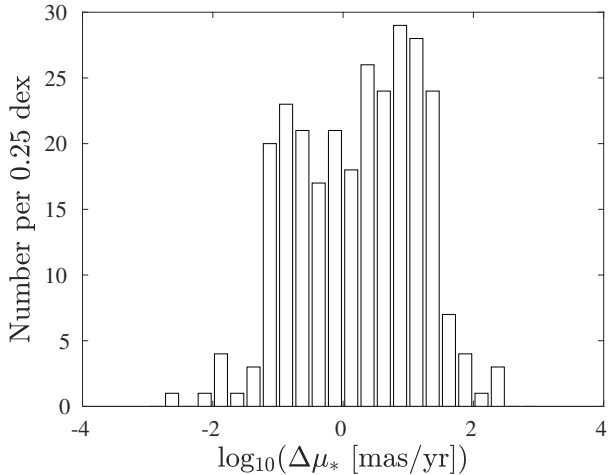}}
\caption{The distribution of the errors in the total proper motion of the 277 pulsars with measured proper motions in the ATNF pulsars catalog.}
\label{fig:psr_pm_error}
\end{figure}

Out of 2636 pulsars in the ATNF catalog, 277 have such measurements.
Figure~\ref{fig:psr_pm_error} presents the distribution of errors in the proper motions of these pulsars.
The typical error in the pulsars proper motion is an order of magnitude
larger than that of GAIA-DR2 stars.
However, these proper motions are good enough to predict the position of a pulsar
to an accuracy of $\sim0.''1$ in the next 100\,yr.
% Figure - proper motion error distribution

For each pulsar, I used the {\tt catsHTM} tool (Soumagnac \& Ofek 2018)
to query for all the GAIA-DR2 (Gaia Collaboration et al. 2016;
Gaia Collaboration et al. 2018) sources within 1000\,arcsec of the pulsar's cataloged position.
Given the pulsar's and GAIA-DR2 sources position and proper motion
I calculated the closest approach between the stars and the pulsar.
For each closest approach with an angular distance below 100\,arcsec,
that takes place between 2000 and 2100, I calculated
the expected astrometric microlensing center-of-light
deflection as a function of time.
I note that, when the lens distance is small, conjunctions with 
a larger impact parameter can induce considerable light deflections.
However, these deflections will typically change over time
scales of decades, making them less attractive for follow up observations.

The light deflection of the center-of-light of the source
(i.e., the more distant object of the two)
was calculated by taking the positions of the two images of the background star
weighted by their respective magnifications,
and taking into account the star and pulsar parallaxes.
If the star parallax is smaller than two
times the parallax error, I set the distance of the star to
10\,kpc.

In the thin screen, point mass, small angle approximation,
the center-of-light of the source light deflection of the images, as measured relative
to the source in the direction of the lens, is:
\begin{equation}
\delta=\frac{\vert m_{+}\vert\theta_{+} + \vert m_{-}\vert\theta_{-}}{\vert m_{+}\vert+\vert m_{-}\vert} - \beta = \frac{\beta/\theta_{E}}{(\beta/\theta_{E})^{2}+2}\theta_{E},
\label{eq:MeanPos}
\end{equation}
where $\beta$ is the angular distance between the lens and source,
$\theta_{+}$ and $\theta_{-}$ are the positions
of the two images\footnote{Due to the point-mass lens assumption, the third image has infinite demagnification.}
of the source (relative to the lens), given by (e.g., Einstein 1936; Liebes 1964; Refsdal 1964; Narayan \& Bartelmann 1996):
\begin{equation}
\theta_{\pm} = \frac{1}{2}(\beta \pm \sqrt{\beta^{2} +4\theta_{E}^{2}}),
\label{eq:theta_pm}
\end{equation}
and $m_{+}$ and $m_{-}$ are the magnifications of the two images
\begin{equation}
m_{\pm} = \Big[1 - \Big(\frac{\theta_{E}}{\theta_{\pm}}\Big)^{4}\Big]^{-1}.
\label{eq:mu_pm}
\end{equation}
Here, $\theta_{E}$ is the angular Einstein radius, in radians, given by
\begin{equation}
%\theta_{E} = \sqrt{\frac{4GM}{c^{2}} \frac{D_{ls}}{D_{l}D_{s}} },
\theta_{E} = \sqrt{\frac{4GM}{c^{2}} \frac{\pi_{{\rm rel}}}{AU} },
\label{eq:ER}
\end{equation}
where $G$ is the gravitational constant, $M$ is the lens mass,
$c$ is the speed of light, $AU$ is the astronomical unit,
and $\pi_{{\rm rel}}$ is the relative parallax in arcseconds
\begin{equation}
\pi_{{\rm rel}} = \frac{\pi_{{\rm l}}\pi_{{\rm s}}}{\pi_{{\rm ls}}},
\label{eq:pi_rel}
\end{equation}
where $\pi_{{\rm l}}$, $\pi_{{\rm s}}$, and $\pi_{{\rm ls}}$
are the parallaxes between the observer and the lens,
the observer and the source,
and the lens and the source, respectively.
Note that when the images separation ($\theta_{+}+\theta_{-}$)
becomes larger than the instrument resolution, the source deflection angle
should not include the demagnified image contribution.
In such a case, 
\begin{equation}
\delta = \theta_{+} - \beta.
\label{eq:sapp}
\end{equation}
However, for practical purposes, there is no difference
between Equation~\ref{eq:MeanPos} and Equation~\ref{eq:sapp}.
For the case of $\beta\gg\theta_{\rm E}$, we can approximate Equation~\ref{eq:MeanPos}
and Equation~\ref{eq:sapp} using
\begin{equation}
\delta \cong \frac{\theta_{E}^{2}}{\beta} = \frac{4GM}{c^{2}} \frac{\pi_{\rm l}\pi_{\rm s}}{\pi_{\rm ls} AU}\frac{1}{\beta}
\label{eq:del_approx}
\end{equation}
Finally, the time scale for the lensing phenomenon is given by $\sim\theta_{\rm E}/\mu_{*}$, where $\mu_{*}$ is the lens-source relative total proper motion.

Next, I selected sources that have a maximum light deflection
above 50\,$\mu$as (relative to an event with no microlensing).
One obvious contamination is that some pulsars may have
companions -- if all proper motion measurements are correct,
such systems will have constant angular separation between
the pulsar and star as a function of time.
In order to avoid selecting pulsars with a companion,
I selected only sources for which the separation varies by
more than $1''$ over the $\pm20$\,yr of the time of closest approach.
The last criterion can be relaxed when cleaner new versions
of the GAIA catalog become available.

The search utilized code available
as part of the {\tt MATLAB} astronomy and astrophysics toolbox\footnote{https://webhome.weizmann.ac.il/home/eofek/matlab/}
(Ofek 2014).

\section{Candidate events}
\label{sec:Cand}

The selection process described in \S\ref{sec:search}
yielded one candidate for astrometric microlensing events,
involving PSR\,J$0846-3533$.
This event will not introduce noticeable magnification of the background source.
However, it still can be detected by measuring the position
of the background source as a function of time and looking
for deviations from the expected linear plus parallax motion.
This motion can be more complicated if the sources are blended
or if it is an astrometric binary.

The details of the event, including the pulsar's parameters,
are listed in Table~\ref{tab:PSR0846}.
This pulsar have no known companions.
Due to the large uncertanty in the pulsar coordinates
and proper motion, the minimum angular separation, 
and expected light deflection, are highly uncertain.
\begin{deluxetable}{llll}
\tablecolumns{3}
\tablewidth{0pt}
%\tabletypesize{\footnotesize}
\tablecaption{PSR\,J$0846-3533$ conjunction parameters}
\tablehead{
\colhead{Parameter}     &
\colhead{Value}        
}
\startdata
Pulsar data & \\
\hline
Name                   & PSR\,J$084606.060-353340.64$  \\
J2000.0 R.A.           & 08:46:06.060 $\pm0.5''$ \\
J2000.0 Dec.           & $-35$:33:40.64 $\pm0.5''$ \\
Proper motion in R.A.  & $93\pm72$\,mas \\
Proper motion in Dec.  & $-15\pm65$\,mas \\
Coordinates epoch      & MJD 48719  \\
Distance (Dispersion Measure)  & 0.54\,kpc \\
Period                 & 1.1161\,s \\
Period derivative      & $1.60\times10^{-15}$ \\
Characteristic age     & $1.1\times10^{7}$\,yr \\
\hline
\hline
Star data & \\
\hline
J2000.0 R.A.             & 08:46:06.291901 $\pm0.15$\,mas \\
J2000.0 Dec.             & $-35$:33:41.34052 $\pm0.18$\,mas \\
Coordinates epoch        & J2015.5 \\
Proper motion in R.A.    & $-3.17\pm0.32$ \\
Proper motion in Dec.    & $2.80\pm0.35$ \\
Parallax                 & $0.10\pm0.23$ \\
R.A./Dec. correlation    & 0.15 \\
Astrometric excess noise & 0.38\,mas \\
Mag G                    & $19.027\pm0.002$ \\
Mag BP                   & $19.806\pm0.048$ \\
Mag RP                   & $18.197\pm0.018$ \\
\hline
\hline
Time of minimum separation    & J2022.9 \\
Minimum angular separation    & $\sim 0.22''$ \\
Estimated Einstein radius     & 4.5\,mas \\
Maximum astrometric deviation & 91$\mu$as 
\enddata
\tablecomments{Due to the large uncertanty in the pulsar coordinates
and proper motion, the minimum angular separation, 
and expected light deflection, are highly uncertain.
Pulsar proper motion are adopted from Zou et al. (2005).
\label{tab:PSR0846}}
\end{deluxetable}

\begin{figure}
\centerline{\includegraphics[width=8cm]{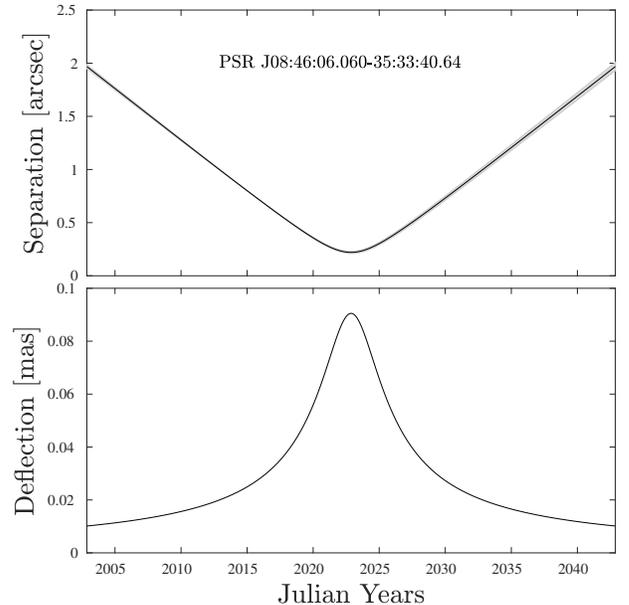}}
\caption{PSR\,J$0846-3533$ angular separation as a function of time from the GAIA star (top panel),
and the predicted center-of-light deflection relative to the source position in the direction of the lens (pulsar). 
The pulsar position is uncertain to about $0.5''$ so the minimum separation is still highly uncertain
and requires verification.}
\label{fig:cand_70_sep_def}
\end{figure}

\begin{figure}
\centerline{\includegraphics[width=8cm]{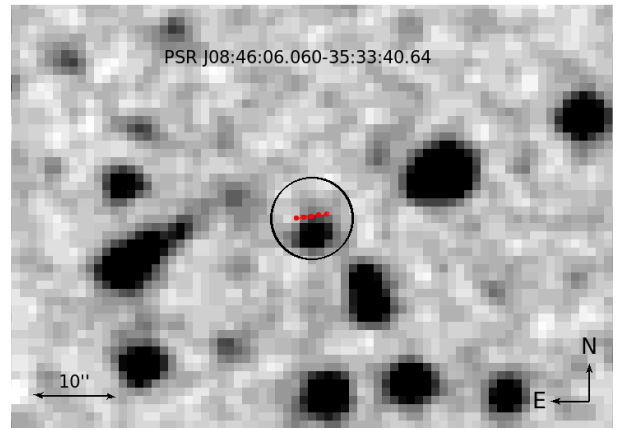}}
\caption{Digitized Sky Survey 2 red image of the field of PSR\,J$0846-3533$.
The image gray-scale is inverted.
The star's position is marked with a black circle.
The red dots mark the pulsar position, every 10 years, from J2002.9 to J2042.9 (East to West).}
\label{fig:J0846m35_dss}
\end{figure}

Figures~\ref{fig:cand_70_sep_def} presents the separation and center-of-light deflection
as a function of time for PSR\,J$0846-3533$.
Finding charts of the pulsar field is shown in
Figures~\ref{fig:J0846m35_dss}.

\section{Discussion}
\label{sec:Disc}

I present a search for conjunctions of pulsars with GAIA-DR2 stars.
I find one such possible event, that need to be verified, in which the astrometric deflection,
from a constant rate motion of the background star,
is expected to be $\sim0.1$\,mas.

This search should be regarded as a preliminary search for candidates.
The pulsar's and star's proper motions and distances should be verified
with future observations, and it will also be useful to obtain
high-resolution imaging of these fields.

An important requirement for such a program is the ability
to measure small astrometric shifts, preferably on the level
of 1--10$\mu$as.
Current state-of-the-art ground-based observations deliver
100\,$\mu$as astrometric precision (e.g., Tendulkar et al. 2012;
Lu et al. 2016).
Therefore, new techniques and methodologies are required
in order to improve the currently available level of precision.
I note that the current limitations likely
result from systematic errors, and therefore, there is a
realistic potential for improvement.

A related important question is how well do we need
to measure the light deflection and distances to the lens
and source in order to measure the lens mass to some accuracy
(see also Gould 2000).
For $\beta\gg\theta_{E}$,
the mass depends linearly on $\beta$, $\delta$ and $1/\pi_{\rm rel}$
(Equation~\ref{eq:del_approx}).
Specifically, the relative error in the mass estimate will be
\begin{equation}
\Big(\frac{\sigma M}{M}\Big)^{2} \propto
\Big(\frac{\sigma \beta}{\beta}\Big)^{2} + 
\Big(\frac{\sigma \delta}{\delta}\Big)^{2} + 
\Big(\frac{\sigma \pi_{\rm rel}}{\pi_{\rm rel}}\Big)^{2}.
\label{eq:MassErr}
\end{equation}
Here, $\sigma{X}$ is the uncertainty in variable $X$.
The relative error in the first term ($\beta$) is expected to
be negligible and the relative errors from the second ($\delta$)
and third ($\pi_{\rm rel}$) terms will likely dominate the errors.
Therefore,
measuring the mass of the NS to better than about 10\% accuracy
will require controlling the systematic errors in the astrometry to better than 10$\mu$as,
and measuring the relative parallax (which depends on $\pi_{\rm l}$ and $\pi_{\rm s}$) to better than 10\%.

Another interesting possibility,
already proposed by Larchenkova \& Doroshenko (1995) and Wex et al. (1996),
is that a pulsar will be lensed by a star.
In principle, this offers the possibility to measure the 
variations in the lensing time delay 
via the timing observations of the pulsar.
The constant time delay adds a constant phase to the pulsar timing,
while the first derivative of the time delay multiplied
by the pulsar period adds a constant
to the pulsar periodicity. Therefore, these terms cannot be measured.
However, the second derivative of the time delay
multiplied by the pulsar period adds a constant
to the first derivative of the pulsar period ($\dot{P}$),
and abrupt variations in $\dot{P}$ (on a year time scale)
can be measured.
The time delay near a point-mass lens potential drops logarithmically
with the impact parameter (see Equation~\ref{eq:Dt1}) and, therefore, the collective
effects of the galactic potential, as well as all the stars near the
line of sight, may be important.
However, when a star is passing with a small impact parameter (e.g., $\ll 1''$)
it will induce variations in the time delay that are relatively abrupt
and that may dominate over all other contributions to the time delay variations.
The relevant formulae for the expected time delay
are provided, along with an example, in Appendix~\ref{app:Shapiro}.
I do~not find good candidates for such events of pulsars lensed
by foreground stars, which are expected to induce variations 
in $\dot{P}$, divided by 1\,s, exceeding $10^{-20}$\,s$^{-1}$.

\acknowledgments

I would like to thank Heng Xu for pointing out to the erroneous declination
of PSR\,J$1856-3754$ in the ATNF catalog.
I am grateful to Andy Gould, Ofer Yaron, and Orly Gnat for comments on the manuscript,
to Boaz Katz for useful discussions, and for the support by
grants from the Israel Science Foundation, Minerva, Israeli Ministry of Technology and Science, the US-Israel Binational Science Foundation,
and the I-CORE Program of the Planning and Budgeting Committee and the Israel Science Foundation.

\appendix
\section{The second derivative of the Shapiro time delay}
\label{app:Shapiro}

Larchenkova \& Doroshenko (1995) and Wex et al.(1996) suggested that pulsar time variations
can be used to detect an unseen mass or measure the mass of stars.

In the limit that the impact parameter is much larger than the Einstein radius,
the geometric time delay can be neglected and the gravitational
time delay is given by the Shapiro time delay
\begin{equation}
\Delta{t} \cong -\sum_{l}{\frac{2GM_{l}}{c^{3}}\ln{[1-cos(\beta_{l})]}}.
\label{eq:Dt1}
\end{equation}
Here, $M_{l}$ is the mass of the $l$-th lens positioned at an angular distance $\beta_{l}$ from the source.

The logarithmic dependence of the time delay on $\beta$
suggests that the stochastic background (e.g., the Galactic potential)
is typically the dominant contributor to the time delay.
However, when a star is passing with a small impact parameter (e.g., $\ll 1''$)
from a pulsar, then the star gravitational potential may dominate the time delay.

For pulsar observations, the Shapiro time delay induces a phase shift to the pulsar observations.
The first derivative of the Shapiro time delay
multiplied by the pulsar period, adds a constant to the measured periodicity.
The second derivative of the Shapiro time delay,
multiplied by the pulsar period, adds a constant to the measured $\dot{P}$.
Therefore, we are interested in the second time derivative of the Shapiro time delay, $\ddot{\Delta{t}}$.
For two sources moving with a relative proper motion $\mu_{*}$ (ignoring parallax),
the angular distance $\beta$ as a function of time $t'$ is
\begin{equation}
\beta=\sqrt{\beta_{m}^{2} +[\mu_{*}(t'-t_{0})]^{2}},
\label{eq:betat}
\end{equation}
where $t_{0}$ is the time of minimum separation, and $\beta_{m}$ is the minimum impact parameter.
Denoting $t=t'-t_{0}$, we get
\begin{equation}
\ddot{\Delta{t}} =  -\frac{GM}{c^{3}} \frac{\mu_{*}^{2} (\beta_{m}^{2} \sin{(\sqrt{t^{2}\mu_{*}^{2} + \beta_{m}^{2}}) - t^{2}\mu_{*}^{2} \sqrt{t^{2}\mu_{*}^{2} + \beta_{m}^{2}}  }  )    }{  (t^{2}\mu_{*}^{2} + \beta_{m}^{2})^{3/2}  (\cos{( \sqrt{t^{2}\mu_{*}^{2} + \beta_{m}^{2}} )} - 1)    }.
\label{eq:Dt2}
\end{equation}

\begin{figure}
\centerline{\includegraphics[width=8cm]{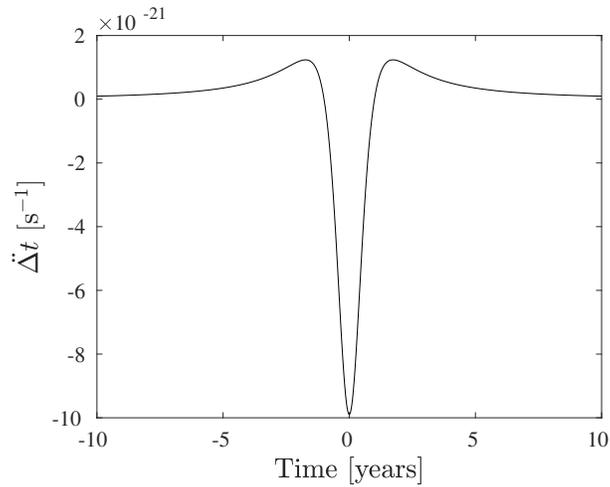}}
\caption{The expected $\ddot{\Delta{t}}$ (neglecting the stochastic background) in the case
of a star passing in the foreground of a pulsar with $\beta_{m}=0.''1$ and $\mu=100$\,mas\,yr$^{-1}$.}
\label{fig:TimeDelay_dt2}
\end{figure}

Figure~\ref{fig:TimeDelay_dt2} shows an example for $\ddot{\Delta{t}}$ as a function of time,
for $\beta_{m}=0.''1$ and $\mu=100$\,mas\,yr$^{-1}$.

%\section{appendix material}
%\label{Ap:}
%\begin{thebibliography}{dummy}
%\bibliographystyle{apj}

%\bibliography{bibliograph.bib}
%\end{thebibliography}

\end{document}